\documentclass[aps,preprint,epsfig,rotate]{revtex4}
\usepackage{graphicx}
\usepackage{bm}
\usepackage{epsfig}

\input epsf
\begin{document}

\title{On hyperfine structure of the rotationally excited 
$1 s_{\mu} n\ell_{e}$-states in helium muonic atoms}

\author{Alexei M. Frolov}
 \email[E--mail address: ]{alex1975frol@gmail.com} 


\affiliation{Department of Applied Mathematics \\
 University of Western Ontario, London, Ontario N6H 5B7, Canada}

\date{\today}

\begin{abstract}
The new approach is develop to determine the hyperfine structure 
of the rotationally excited $1 s_{\mu} n \ell_{e}$-states in the 
three-body helium muonic ${}^{3}$He $\mu \; e$ and ${}^{4}$He 
$\mu \; e$ atoms. The formulas derived in this study allow one 
to evaluate the hyperfine structure splittings for an arbitrary 
rotationally excited (bound) $1 s_{\mu} n \ell_{e}$-state in 
the three-body helium muonic atoms. Similar formulas have never 
been produced in earlier studies and this explains why these 
hyperfine structure splittings are still unknown for these bound 
states.  

\end{abstract}

\maketitle

\newpage

\section{Introduction}

In this short communication we derive the formulas which can be used to determine 
the hyperfine structure of the rotationally excited $1 s_{\mu} n \ell_{e}$-states 
in the helium muonic ${}^{3}$He $\mu \; e$ and ${}^{4}$He $\mu \; e$ atoms. As is 
well known the hyperfine structure of any one-electron atomic and/or quasi-atomic 
system can be analyzed by using the relativistic Dirac equation \cite{Dir28}. For 
one-electron atoms and ions this was done by E. Fermi in his works published in 
1930 \cite{FermiL1}. Later the same methods were generalized to various atoms and 
ions (see, e.g., \cite{Sobel}, \cite{Heine} and references therein). In late 
1970's similar research of the hyperfine structure started (see, e.g., 
\cite{Hughes} - \cite{Fro2}) for the helium muonic atoms. By the helium muonic 
atom we mean the three-body, quasi-atomic (neutral) systems which contains one 
electron $e^{-}$ and two heavy particles: the `central' helium nucleus He$^{2+}$ 
and negatively charged muon $\mu^{-}$. In fact, between 1978 and 1981 the two 
helium atoms ${}^{3}$He $\mu \; e$ and ${}^{4}$He $\mu \; e$ were created 
experimentally and investigated. Discussion of early experiments and many useful 
references can be found in \cite{Hughes} and \cite{Huan1}. 

In our previous paper \cite{Fro2026} we investigated the hyperfine structure of 
the three-body helium-muonic ${}^{3}$He $\mu \; e$ and ${}^{4}$He $\mu \; e$ 
atoms in the bound $S(L = 0)-$states which are designated below as the bound 
$1 s_{\mu} n s_{e}$-states. The general formula for the Hamiltonian of hyperfine 
structure in the case of bound $1 s_{\mu} n s_{e}$-states written in atomic 
units, where $\hbar = 1, m_e = 1, e = 1$ (and also $4 \pi \epsilon_0 = 1$), 
takes the form (see, e.g., \cite{Fro2026}):  
\begin{eqnarray}
 \hat{H}_{HFS} &=& - \frac{8 \pi \alpha^{2}}{3} \; g_{N} g_{\mu} \; 
 \Bigl(\frac{m_e}{M_p}\Bigr) \; \Bigl(\frac{m_e}{M_{\mu}}\Bigr) \; (\mu_{B} 
 c)^{2} \; \langle \delta({\bf r}_{N \mu}) \rangle \; ({\bf I}_{N} \cdot 
 {\bf s}_{\mu}) \nonumber \\
  &-&\frac{8 \pi \alpha^{2}}{3} \; g_{e} g_{\mu} \; 
  \Bigl(\frac{m_e}{M_{\mu}}\Bigr) \; (\mu_{B} c)^{2} \; \langle \delta({\bf 
  r}_{\mu e}) \rangle \; ({\bf s}_{\mu} \cdot {\bf s}_{e}) \; \;  \nonumber \\
  &-&\frac{8 \pi \alpha^{2}}{3} g_{N} g_{e} \; \Bigl(\frac{m_e}{M_p}\Bigr) \; 
  (\mu_{B} c)^{2} \; \langle \delta({\bf r}_{N e}) \rangle ({\bf I}_{N} \cdot 
  {\bf s}_{e}) \; \; , \; \; \label{HhfsA1} 
\end{eqnarray}
where $\alpha = 7.2973525693 \cdot 10^{-3}$ is the dimensionless fine structure 
constant, $c$ is the speed of light in vacuum, $\mu_B = \frac{\hbar e}{2 c m_e}$ 
is the Bohr magneton, while $\mu_B = \frac{\hbar e}{2 c M_P} = \mu_B \Bigl( 
\frac{m_e}{M_p} \Bigr)$ is the nuclear magneton, where $M_p$ = 1836.15267343 
$m_e$ is the proton mass. The notation $m_e$ designates the electron 
mass-at-rest. The actual masses of the helium atomic nuclei are  5495.88528007 
$m_e$ (${}^{3}$He) and 7294.29954142 $m_e$ (${}^{4}$He), while the muon 
mass-at-rest is 206.7682830 $m_e$. Everywhere below in this study the muon mass 
is designated as $M_{\mu}$. Furthermore, the numerical values of all physical 
constants and properties of particles have been chosen from \cite{NIST}. Also 
in Eq.(\ref{HhfsA1}) the notation $\langle \delta( {\bf}_{i j} ) \rangle$ stands 
for the expectation value of the delta-function between particles $i$ and $j$ 
(see below), which must be taken in atomic units, where $\hbar = 1, m_e = 1, 
\mid e \mid = 1$ and $4 \pi \epsilon_0 = 1$. In atomic units the speed of light 
$c$ equals $\alpha^{-1}$ = 137.0359991. Note that in atomic units one finds $c 
= \alpha^{-1}, \mu_{B} = \frac{\alpha}{2}$ and $\mu_{B} c = \frac12$. This 
transforms Eq.(\ref{HhfsA1}) to the final formula for the $\hat{H}_{HFS}$ 
Hamiltonian:  
\begin{eqnarray}
  \hat{H}_{HFS} &=& - \frac{2 \pi \alpha^{2}}{3} \; g_{N} g_{\mu} \; 
  \Bigl(\frac{m_e}{M_p}\Bigr) \; \Bigl(\frac{m_e}{M_{\mu}}\Bigr) \; \langle 
  \delta({\bf r}_{N \mu}) \rangle \; ({\bf I}_{N} \cdot {\bf s}_{\mu}) 
  \nonumber \\
 &-& \frac{2 \pi \alpha^{2}}{3} \; g_{e} g_{\mu} \; 
 \Bigl(\frac{m_e}{M_{\mu}}\Bigr) \; \langle \delta({\bf r}_{\mu e}) \rangle 
 \; ({\bf s}_{\mu} \cdot {\bf s}_{e}) \;   
 - \; \frac{2 \pi \alpha^{2}}{3} g_{N} g_{e} \; \Bigl(\frac{m_e}{M_p}\Bigr) 
 \langle \delta({\bf r}_{N e}) \rangle ({\bf I}_{N} \cdot {\bf s}_{e}) \; 
 \; , \; \; \label{HhfsA} 
\end{eqnarray}
which is used for the direct numerical calculations of the hyperfine structure 
splittings in all bound $S(L = 0)-$states of the helium-muonic atoms (see, e.g., 
\cite{Fro2026}. 

In addition to this, we have to note that the notation $N$ in Eq.(\ref{HhfsA1})  
and Eq.(\ref{HhfsA}) stands for the ${}^{3}$He and ${}^{4}$He nuclei, while 
symbols $e$ and $\mu$ mean the electron and muon, respectively. The $g_{e}, 
g_{\mu}$ and $g_{N}$ the electron, muon and helicon spin $g-$factors, which are 
the dimensionless ratios of the electron, muon and nucleus magnetic momenta to 
their intrinsic spins, respectively. As mentioned above the notation $\langle 
\delta({\bf r}_{i j}) \rangle$ traditionally denotes the expectation value of 
the $(ij)$ interparticle (or two-particle) delta-function, where $i \ne j = (e, 
\mu, N)$. Each of these expectation values are determined by using spatial 
parts of the bound state wave functions $\Psi$, i.e., the unit-norm wave 
functions of the corresponding (bound) states. The same rule is applied to any 
formula which appears below and contains the expectation values. 

Also, in Eq.(\ref{HhfsA}) and other similar equations below the symbols ${\bf 
I}_{N}, {\bf s}_{\mu}$ and ${\bf s}_{e}$ designate the spins of the helium 
nucleus, muon and electron, respectively. The absolute values of these three 
spins are always either integer, or semi-integer. For the helium-4 muonic atom 
we have ${\bf I}_N = 0$ and our formula, Eq.(\ref{HhfsA}), drastically 
simplifies, since it is written in the one-term form:
\begin{eqnarray}
  & & H_{HFS} = - \frac{2 \pi \alpha^{2}}{3} \; g_{e} g_{\mu} \; 
  \Bigl(\frac{m_e}{M_{\mu}}\Bigr) \; \langle \delta({\bf r}_{\mu e}) \rangle 
  \; ({\bf s}_{\mu} \cdot {\bf s}_{e}) \; . \; \label{HhfsB} 
\end{eqnarray} 
In general, for the bound $S(L = 0)-$states the Hamiltonians of hyperfine 
structure, Eqs.(\ref{HhfsA}) and/or (\ref{HhfsB}), are the matrices with the 
dimension $(2 I + 1) (2 s_{\mu} + 1) (2 s_{e} + 1)$. To determine the hyperfine 
structure of such a helium muonic atoms we need to determine all eigenvalues 
(with the corresponding multiplicities) of this matrix. The energy differences 
between eigenvalues (or groups of eigenvalues) give us the corresponding 
hyperfine structure splittings. In general, in atomic and muon-atomic physics 
such energy differences are considered as some `splittings', if they are 
expressed in $MHz$. The whole procedure is rigorously described in 
\cite{Fro2026} and here we do not want to repeat its description in detail. 

The formulas, Eq.(\ref{HhfsA1}), Eq.(\ref{HhfsA1}) and Eq.(\ref{HhfsA}), allow 
one to determine the hyperfine structure splittings for the bound $1 s_{\mu} n 
s_{e}$-states in the helium muonic ${}^{3}$He $\mu \; e$ and ${}^{4}$He $\mu \; 
e$ atoms. In particular, in our paper \cite{Fro2026} the hyperfine structure 
splittings $\Delta$ for some bound $1 s_{\mu} n s_{e}$-states, where $n = 1, 2, 
3$, have been evaluated (in $MHz$) as:
\begin{eqnarray}
 \; \; 4166.38685445(3) \; ({\rm for} \; n = 1) \; , \; \; 520.78609177(5) \; 
 ({\rm for} \; n = 2) \; , \; \; 154.30584687(5) \; ({\rm for} \; n = 3) \; 
 \nonumber 
\end{eqnarray}
for the helium-muonic ${}^{3}$He $\mu \; e$ atom and 
\begin{eqnarray}
 \; \;  4464.55409995(3) \; ({\rm for} \; n = 1) \; , \; \; 558.05491883(5) 
 \; ({\rm for} \; n = 2) \; , \; \; 165.34833511(5) \; ({\rm for} \; n = 3) 
 \; \nonumber 
\end{eqnarray}
for the helium-muonic ${}^{4}$He $\mu \; e$ atom. In reality, the overall 
numerical accuracy of current variational computations for the bound $1 s_{\mu} 
n s_{e}$-states in three-body helium muonic atoms is very high and constantly 
increases. Briefly, based on these results we can say that the long standing 
problem of the hyperfine structure splittings for the bound $1 s_{\mu} n 
s_{e}$-states in helium muonic atoms has been solved completely and accurately.
It is clear that there are quite a few small additional corrections, e.g., the 
lowest-order relativistic and QED corrections, which must be taken into account. 
But here we cannot discuss them. Instead, our next goal is to investigate the 
hyperfine structure of the rotationally excited states in the both helium muonic 
${}^{3}$He $\mu \; e$ and ${}^{4}$He $\mu \; e$ atoms. This problem is analyzed 
in the following Sections. 

\section{Hyperfine structure of the rotationally excited states} 

The next step in research of the hyperfine structure of the helium muonic 
atoms is to analyze a similar problem for the rotationally excited $1 s_{\mu} 
n \ell_{e}$-states in these quasi-atomic systems. Right now, it is not an easy 
task, since currently we do not have any analytical formula(s) which can be 
used to determine the hyperfine structure splittings in the rotationally 
excited, bound states of the helium muonic atoms. The main goal of this 
Section is to derive the closed analytical formulas which allow one to 
evaluate the hyperfine structure splittings for arbitrary rotationally excited 
$1 s_{\mu} n \ell_{e}$-state(s) in the helium muonic atoms. Currently, we know 
that many of these rotationally excited $1 s_{\mu} n \ell_{e}$-states in the 
helium-muonic atoms are stable (see, discussion and our results in 
\cite{Fro2026}). In particular, the rotationally excited $1 s_{\mu} 2 p_{e}$- 
and $1 s_{\mu} 3 d_{e}$-states are stable and they are the lowest-by-energy 
bound P- and D-states in the helium muonic ${}^{3}$He $\mu \; e$ and 
${}^{4}$He $\mu \; e$ atoms atoms.  

First of all, let us derive the general formulas which are needed to determine 
the hyperfine structure of the rotationally excited $1 s_{\mu} n \ell_{e}$
states in the helium muonic ${}^{3}$He $\mu \; e$ and ${}^{4}$He $\mu \; e$ 
atoms. For this purpose let us consider the total vector potential $\vec{A}$ 
and magnetic field strength $\vec{H}$ which act on a single electron which 
moves in the magnetic field created by the central nucleus $N$ and negatively 
charged muon $\mu^{-}$. The vector potential $\vec{A}$, which acts on such an 
electron $e^{-}$, is written in the form (see, e.g., \cite{Jackson}, 
\cite{LLQ}):
\begin{eqnarray} 
 \vec{A} = \frac{( \vec{\mu}_{N} \times \vec{r}_{31} )}{r^{3}_{31}} + 
 \frac{( \vec{\mu}_{\mu} \times \vec{r}_{21} )}{r^{3}_{21}} \; \; \; , \; 
 \; \label{A1}
\end{eqnarray} 
where the particles 3 and 2 mean the atomic (helium) nucleus and negatively 
charged muon $\mu^{-}$, while the particle 1 always designates the electron 
$e^{-}$. The notations $\vec{\mu}_{N}$ and $\vec{\mu}_{\mu}$ stand for the 
magnetic moments of the nucleus and $\mu^{-}$ muon, respectively. This 
system of notation is used everywhere below in this study. As follows from 
these formulas the magnetic interaction between bound (atomic) electron and 
total vector potential equals to the scalar product of the vector potential 
$\vec{A}$ and electron's momentum $\vec{p}_e$
\begin{eqnarray} 
 \frac{e \hbar}{m_e c} \; \vec{A} \cdot \vec{p}_e =  \frac{e \hbar}{m_e c} 
 \; \Bigl[ \frac{1}{r^{3}_{31}} \; ( \vec{\mu}_{N} \cdot \vec{\ell}_{31} ) 
 + \frac{1}{r^{3}_{21}} \; ( \vec{\mu}_{\mu} \cdot \vec{\ell}_{21} ) 
 \Bigr] \; \; \; . \; \; \label{Apt}
\end{eqnarray} 
where $e = \mid e^{-} \mid \ge 0$ is the absolute electric charge of the 
electron $e^{-}$. The mixed product included in the last equation is 
transformed in the following way: $( \vec{\mu}_{k} \times \vec{r}_{k 1} ) 
\cdot \vec{p}_{e} =  \vec{\mu}_{k} \cdot ( \vec{r}_{k 1} \times \vec{p}_{e} 
) = \vec{\mu}_{k} \cdot \vec{\ell}_{k 1}$, where $k = (N, \mu) = (3, 2)$ and 
$\vec{\ell}_{k 1}$ is the electron's angular moment in respect to the 
`central' particle $k$. In our analysis below we shall follow \cite{Jackson}, 
\cite{LLQ}. 

The total magnetic field strength $\vec{H}$, which acts on the same electron, 
is represented as the following sum of two terms:
\begin{eqnarray} 
 \vec{H} = \frac{ 3 \; \vec{n}_{31} \; ( \vec{\mu}_{N} \cdot \vec{n}_{31} ) 
 - \vec{\mu}_{N} }{r^{3}_{31}} + \frac{ 3 \; \vec{n}_{21} \; ( 
 \vec{\mu}_{\mu} \cdot \vec{n}_{21}) - \vec{\mu}_{\mu}}{r^{3}_{21}} \; \; 
 \; , \; \; \label{H1}
\end{eqnarray} 
where $\vec{n}_{31} = \frac{\vec{r}_{31}}{r_{31}}$ and $\vec{n}_{21} = 
\frac{\vec{r}_{21}}{r_{21}}$ are the two unit vectors oriented into the 
$\vec{r}_{31}$ and $\vec{r}_{21}$ directions, respectively. The same notations
are used everywhere below in this study. Also, the first term in this equation 
represents the nucleus magnetic component of the total magnetic field strength 
$\vec{H}$, while the second term describes similar contribution from the muon 
magnetic component. The corresponding part of the magnetic interaction between 
these two heavy particles and electron equals to the scalar product 
$\vec{\mu}_e \cdot \vec{H}$, where $\vec{\mu}_e$ is the magnetic moment of the 
electron and $\vec{H}$ is defined in Eq.(\ref{H1}). Now, we can introduce the 
Hamiltonian of hyperfine structure (see, e.g., \cite{Jackson}) for the 
electron-nucleus pair:  
\begin{eqnarray} 
 {\cal H}^{(N e)}_{HFS} = - \frac{8 \pi}{3} \vec{\mu}_{N} \cdot \vec{\mu}_{e} \; 
 \delta(\vec{r}_{31}) - \frac{1}{r^{3}_{31}} \; \Bigl[ 3 \; (\vec{n}_{31} \cdot 
 \vec{\mu}_{N}) \; ( \vec{n}_{31} \cdot \vec{\mu}_{e} ) - ( \vec{\mu}_{N} \cdot 
 \vec{\mu}_{e} ) + \frac{e}{m_e c} \; ( \vec{\ell}_{31} \cdot \vec{\mu}_{N} ) 
 \Bigr] \; \; \; . \; \; \label{HNe} 
\end{eqnarray} 
Analogously, we can write the Hamiltonian of hyperfine structure for the 
electron-muon pair 
\begin{eqnarray} 
 {\cal H}^{(\mu e)}_{HFS} = - \frac{8 \pi}{3} \vec{\mu}_{\mu} \cdot 
 \vec{\mu}_{e} \; \delta(\vec{r}_{21}) - \frac{1}{r^{3}_{21}} \; \Bigl[ 3 \; 
 (\vec{n}_{21} \cdot \vec{\mu}_{\mu}) \; ( \vec{n}_{21} \cdot \vec{\mu}_{e} ) 
 - ( \vec{\mu}_{\mu} \cdot \vec{\mu}_{e} ) + \frac{e}{m_e c} \; ( 
 \vec{\ell}_{21} \cdot \vec{\mu}_{\mu} ) \Bigr] \; \; \; \; \label{Hmue} 
\end{eqnarray} 
and similar Hamiltonian for the muon-nucleus pair 
\begin{eqnarray} 
 {\cal H}^{(N \mu)}_{HFS} = - \frac{8 \pi}{3} \vec{\mu}_{N} \cdot 
 \vec{\mu}_{\mu} \; \delta(\vec{r}_{32}) - \frac{1}{r^{3}_{32}} \; \Bigl[ 3 
 \; (\vec{n}_{32} \cdot \vec{\mu}_{N}) \; ( \vec{n}_{32} \cdot \vec{\mu}_{\mu} 
 ) - ( \vec{\mu}_{N} \cdot \vec{\mu}_{\mu} ) + \frac{e}{M_{\mu} c} \; ( 
 \vec{\ell}_{32} \cdot \vec{\mu}_{N} ) \Bigr] \; \; \; . \; \; \label{HNmu} 
\end{eqnarray} 

The total hyperfine Hamiltonian is written as the sum of these three terms:  
\begin{eqnarray} 
 {\cal H}_{HFS} = {\cal H}^{(N e)}_{HFS} + {\cal H}^{(\mu e)}_{HFS} + 
 {\cal H}^{(N \mu)}_{HFS} \; \; \; . \;  \; \label{Htotal} 
\end{eqnarray} 
In actual applications to the bound states in helium muonic atoms one usually 
finds that the total Hamiltonian, Eq.(\ref{Htotal}), contains a number of terms 
which either equal zero identically, or provide only a very small (negligible) 
contribution. It is clear that similar terms must be excluded from the 
Hamiltonian, Eq.(\ref{Htotal}), of the hyperfine structure. In reality, this 
can be done by considering the internal structure of the three-body helium 
muonic atoms which is discussed in the Appendix. 

Now, let us consider the electron-nucleus part of our hyperfine Hamiltonian 
${\cal H}^{(N e)}_{HFS}$, Eq.(\ref{HNe}). In atomic units this term takes the 
from: 
\begin{eqnarray} 
  {\cal H}^{(N e)}_{HFS}(1)  = - \frac{8 \pi}{3} \vec{\mu}_{N} \cdot 
  \vec{\mu}_{e} \; \langle \delta({\bf r}_{N e}) \rangle \; &=& - \frac{8 
  \pi}{3} \; (c \mu_{B} )^{2} \; \frac{g_{e} g_N}{M_{p}} \; ({\bf I}_{N} 
  \cdot {\bf s}_{e}) \; \langle \delta({\bf r}_{N e}) \rangle \nonumber \\ 
  &=& - \frac{2 \pi \alpha^{2}}{3} \; \frac{g_{e} g_N}{M_{p}} \; ({\bf 
  I}_{N} \cdot {\bf s}_{e}) \; \; \langle \delta({\bf r}_{N e}) \rangle 
  \; \; , \; \;  \label{Term1A1} 
\end{eqnarray} 
where $\mu_{B} = \frac{e \hbar}{2 m_e c}$ is the Bohr magneton, which in 
atomic units equals $\frac12 \alpha$, and $M_p = \frac{M_p}{m_e}$ is the 
dimensionless ratio of the proton and electron masses. Note that in 
Eq.(\ref{Term1A1}) and other similar equations below the expectation values 
must be determined by using the bound state wave functions of the three-body 
helium muonic atoms. In Eq.(\ref{Term1A1}) and other equation below all 
vectors associated with actual particles are designated by the bold letters. 

The second electron-nucleus term is written in a tensor form 
\begin{eqnarray} 
  {\cal H}^{(N e)}_{HFS}(2) = - \frac14 \; \alpha^{2} \; \frac{g_{e} 
  g_N}{M_{p}} \; \Bigl[ 3 \; \langle \; \frac{1}{r^{3}_{31}} \; 
  ({\bf I}_{N} \cdot {\bf n}_{31}) \; ({\bf n}_{31} \cdot {\bf s}_{e}) 
  \; \rangle - \langle \frac{1}{r^{3}_{31}} \rangle \; ({\bf I}_{N} 
  \cdot {\bf s}_{e}) \Bigr] \; . \; \label{Term1A2} 
\end{eqnarray} 
The last (or third) term included in the electron-nucleus part of the 
hyperfine Hamiltonian is the spin-orbital term 
\begin{eqnarray} 
  {\cal H}^{(N e)}_{HFS}(3) = - \langle \frac{1}{r^{3}_{21}} \rangle \; 
  \Bigl(\frac{e \; \hbar}{m_e c}\Bigr) \; \vec{\mu}_{N} \cdot 
  \vec{\ell}_{31} &=& - \frac12 \; \alpha^{2} \; \mu^{2}_B \; 
  \frac{g_N}{M_{p}} \; \langle \frac{1}{r^{3}_{31}} \rangle \; 
  ({\bf I}_{N} \cdot {\bf L}_{31}) \; \nonumber \\
  &=& - \frac12 \alpha^{2} \; \frac{g_N}{M_{p}} \; \langle 
  \frac{1}{r^{3}_{31}} \rangle \; ({\bf I}_{N} \cdot {\bf L}_{31}) \; , 
  \; \label{Term1A3} 
\end{eqnarray} 
where $\frac{e \; \hbar}{m_e \; c} = 2 \; \mu_B \; \alpha$ and ${\bf L}_{31} 
= \vec{\ell}_{31}$ is the electron's angular momentum in respect to the 
helium nucleus. This term describes interaction between the nuclear spin and 
orbital motion of the electron. 

The formula for the three-term electron-muon part in Eq.(\ref{Htotal}) of the 
hyperfine Hamiltonian ${\cal H}^{(\mu e)}_{HFS} = {\cal H}^{(\mu e)}_{HFS}(1) 
+ {\cal H}^{(\mu e)}_{HFS}(2) + {\cal H}^{(\mu e)}_{HFS}(3)$ is derived 
analogously. In particular, in this part one finds the explicit expression for 
the first spin-spin term:   
\begin{eqnarray} 
  {\cal H}^{(\mu e)}_{HFS}(1) = - \frac{2 \pi \alpha^{2}}{3} \; \frac{g_{e} 
  g_{\mu}}{M_{\mu}} \; ({\bf s}_{\mu} \cdot {\bf s}_{e}) \; \; \langle 
  \delta({\bf r}_{\mu e}) \rangle \; \; , \; \;  \label{Term2B1} 
\end{eqnarray} 
where $M_{\mu} = \frac{M_{\mu}}{m_e}$ is the dimensionless ratio of the muon 
and electron masses. The second electron-muon spin-spin term is also written 
in a tensor form: 
\begin{eqnarray} 
 {\cal H}^{(\mu e)}_{HFS}(2) = - \frac14 \; \alpha^{2} \; \frac{g_{e} 
 g_{\mu}}{M_{\mu}} \; \Bigl[ 3 \; \langle \; \frac{1}{r^{3}_{21}} \; 
 ({\bf s}_{\mu} \cdot {\bf n}_{21}) \; ({\bf n}_{21} \cdot {\bf s}_{e}) \; 
 \rangle - \langle \frac{1}{r^{3}_{21}} \rangle \; ({\bf s}_{\mu} \cdot 
 {\bf s}_{e}) \; \Bigr] \; . \; \label{Term22B} 
\end{eqnarray} 
The third (or spin-orbital) term in the electron-muon part of the Hamiltonian 
of hyperfine structure ${\cal H}^{(\mu e)}_{HFS}(3)$ is  
\begin{eqnarray} 
 {\cal H}^{(\mu e)}_{HFS}(3) = - \Bigl(\frac{e \; \hbar}{m_e c}\Bigr) 
 \; \mu_B \; \frac{g_{\mu}}{M_{\mu}} \; \langle \frac{1}{r^{3}_{21}} \rangle 
 \; ({\bf s}_{\mu} \cdot {\bf L}_{21}) = - \frac12 \; \alpha^{2} \; 
 \frac{g_{\mu}}{M_{\mu}} \; \langle \frac{1}{r^{3}_{21}} \rangle \; 
 ({\bf s}_{\mu} \cdot {\bf L}_{21}) \; , \; \label{Term23} 
\end{eqnarray} 
where $\mu_{B} = \frac{e \hbar}{2 m_e c} = \frac12 \alpha$ in atomic units 
and ${\bf L}_{21} = \vec{\ell}_{21}$ is the electron's angular momentum in 
respect to the negatively charged muon $\mu^{-}$. This term describes the 
direct interaction between the muon spin ${\bf s}_{\mu}$ and orbital motion 
of the electron in respect to the negatively charged muon $\mu^{-}$. 

The muon-nucleus part of the hyperfine Hamiltonian ${\cal H}^{(N 
\mu)}_{HFS}$, Eq.(\ref{HNmu}), in the helium muonic atoms contains only 
one term which takes  a familiar form:  
\begin{eqnarray} 
 {\cal H}^{(N \mu)}_{HFS}(1) = - \frac{2 \pi \alpha^{2}}{3} \; g_{N} 
 g_{\mu} \; \Bigl(\frac{m_e}{M_p}\Bigr) \; \Bigl(\frac{m_e}{M_{\mu}}\Bigr) 
 \; \langle \delta({\bf r}_{N \mu}) \rangle \; ({\bf I}_{N} \cdot 
 {\bf s}_{\mu}) \; . \; \label{Term3C3} 
\end{eqnarray}
The total hyperfine Hamiltonian of the helium muonic atoms is the sum of 
three terms (or three partial Hamiltonians) mentioned above, i.e., ${\cal 
H}^{(N e)}_{HFS}, {\cal H}^{(\mu e)}_{HFS}$ and ${\cal H}^{(N \mu)}_{HFS}$. 
However, some of the expectation values equal zero identically and they do 
not contribute into the final sum. For instance, in the $^{4}$He $\mu \; 
e$ atom all terms which include the ${\bf I}_{N}$ `vector' have zero 
contribution, since $\mid {\bf I}_{N} \mid = 0$ for the $^{4}$He nucleus. 
This problem is considered below (see, also Appendix). Note that each of 
these partial Hamiltonians ${\cal H}^{(N e)}_{HFS}, {\cal H}^{(\mu 
e)}_{HFS}$ and ${\cal H}^{(N \mu)}_{HFS}$ is a matrix (or operator) which 
depend upon the spin and/or components of angular momenta only. All spatial 
coordinates of three particles are included only in the expectation values 
which are the contributing parts of these partial Hamiltonians.  

Note that as follows from the formulas derived in this Section almost all 
of them contain either the radial expectation values $\langle 
\frac{1}{r^{3}_{31}} \rangle$ and $\langle \frac{1}{r^{3}_{21}} \rangle$, 
or their tensor analogues. Let us explain briefly the process of 
calculations of these expectation values in the hydrogen-like atoms and 
ions. First of all, it is clear that these expectation values are not 
singular for the rotationally excited states. Second, the explicit 
formula for the electron-nucleus expectation values $\langle 
\frac{1}{r^{3}_{31}} \rangle$ determined on the pure hydrogenic $\mid n, 
\ell, m \rangle $ states in one-electron atom/ion with the nuclear charge
$Q e$ is written in the form (see, e.g., \cite{Sobel}) 
\begin{eqnarray} 
  \langle n, \ell, m \mid \Bigl(\frac{Q e^{2}}{a_{0}}\Bigr)^{3} \cdot 
  \Bigl(\frac{a_{0}}{r^{3}_{31}}\Bigr)^{3} \mid n, \ell, m 
  \rangle = \Bigl( Q \; Ry \Bigr)^{3} \; \frac{1}{n^{3} \ell (\ell + 1) 
  ( \ell + \frac12\Bigr)} \; \; \; , \; \; \; \label{r31*3} 
\end{eqnarray}
which does not depend upon the `magnetic' quantum number $m$ (it could 
be expected). In Eq.(\ref{r31*3}) the atomic electron-nucleus distance 
$r_{31}$ must be expressed in the Bohr radius $a_0$, where $a_0 = 
\frac{\hbar^{2}}{m_e \; e^{2}}$, while $Ry = \frac{m_e e^{4}}{\hbar^2}$ 
is one Rydberg, which is an important constant in atomic physics. In 
atomic units we always have $a_0 = 1, Ry = 1$ and Eq.(\ref{r31*3}) takes 
the form
\begin{eqnarray} 
  \langle n, \ell, m \mid \frac{Q^{3}}{r^{3}_{31}} \mid n, \ell, m 
  \rangle = \frac{Q^{3}}{n^{3} \ell (\ell + 1) ( \ell + \frac12\Bigr)} 
  \; \; \; , \; \; \; \label{r31*31} 
\end{eqnarray}
where $Q$ is the electric charge of atomic nucleus, i.e., $Q$ is a 
positive, integer number. Now, it is easy to see that for positive 
integer $\ell$ ($\ell \ge 1$) the expression in Eq.(\ref{r31*31}) is 
not singular. For the electron-nucleus term in the helium muonic 
${}^{3}$He $\mu \; e$ and ${}^{4}$He $\mu \; e$ atoms one finds in 
Eq.(\ref{r31*31}) that $Q = 1$.

\section{Total hyperfine Hamiltonian of the helium-muonic atoms}

The total Hamiltonian of hyperfine structure in any helium muonic atom, which 
is designated below as $\hat{H}_{HFS}$, is the sum of all partial 
$HFS-$Hamiltonians ${\cal H}^{(N e)}_{HFS}, {\cal H}^{(\mu e)}_{HFS}$ and 
${\cal H}^{(N \mu)}_{HFS}$ mentioned above in Eqs.(\ref{Term1A1}) - 
(\ref{Term3C3}). This total Hamiltonian of the hyperfine structure is written 
in the form
\begin{eqnarray} 
  \hat{H}_{HFS} &=& - \frac{2 \pi \alpha^{2}}{3} \; \Bigl\{ \frac{g_{e} 
  g_N}{M_{p}} \; \; \langle \delta({\bf r}_{N e}) \rangle \; ({\bf I}_{N} 
  \cdot {\bf s}_{e}) + \; \frac{g_{e} g_{\mu}}{M_{\mu}} \; \; \langle 
  \delta({\bf r}_{\mu e}) \rangle \; ({\bf s}_{\mu} \cdot {\bf s}_{e}) \; 
  \nonumber \\
 &+& g_{N} g_{\mu} \; \Bigl(\frac{m_e}{M_p}\Bigr) \; 
 \Bigl(\frac{m_e}{M_{\mu}}\Bigr) \; \langle \delta({\bf r}_{N \mu}) \rangle 
 \; ({\bf I}_{N} \cdot {\bf s}_{\mu}) \Bigr\} \nonumber \\ 
 &-& \frac14 \; \alpha^{2} \; \Bigl\{ \frac{g_{e} g_N}{M_{p}} \; \Bigl[ 3 
 \; \langle \; \frac{1}{r^{3}_{31}} \; ({\bf I}_{N} \cdot {\bf n}_{31}) \; 
 ({\bf n}_{31} \cdot {\bf s}_{e}) \; \rangle \; - \; \langle 
 \frac{1}{r^{3}_{31}} \rangle \; ({\bf I}_{N} \cdot {\bf s}_{e}) \Bigr] 
 \nonumber \\ 
 &+& \frac{g_{e} g_{\mu}}{M_{\mu}} \; \Bigl[  3 \; \langle \; 
 \frac{1}{r^{3}_{21}} \; ({\bf s}_{\mu} \cdot {\bf n}_{21}) \; 
 ({\bf n}_{21} \cdot {\bf s}_{e}) \; \rangle \; - \langle 
 \frac{1}{r^{3}_{21}} \rangle \; ({\bf s}_{\mu} \cdot {\bf s}_{e}) \; 
 \Bigr] \Bigr\} \; \nonumber \\ 
 &-& \frac12 \alpha^{2} \; \Bigl[ \frac{g_N}{M_{p}} \; \langle 
 \frac{1}{r^{3}_{31}} \rangle \; ({\bf I}_{N} \cdot {\bf L}_{31}) + 
 \frac{g_{\mu}}{M_{\mu}} \; \langle \frac{1}{r^{3}_{21}} \rangle \; 
 ({\bf s}_{\mu} \cdot {\bf L}_{21}) \; \Bigr] \; \; \; \; , 
 \; \label{Htot} 
\end{eqnarray}
where the first term essentially coincides with the Hamiltonian $\hat{H}_{HFS}$ 
defined in Eq.(\ref{HhfsA}). The second term in this equation is a tensor which 
has zero contribution for all bound $S(L = 0)-$states (or for the bound $1 
s_{\mu} n s_{e}-$states) in the helium muonic atoms. The last (or third) term 
is the spin-orbital term which describes interaction between the nuclear 
and/or muonic spins and electron which has non-zero angular momentum in respect 
to each of these particles. As expected the last term in Eq.(\ref{Htot}) does 
not include the electron's $g_e-$factor. This term really contributes only for 
the rotationally excited $1 s_{\mu} n \ell_{e}-$states, while for the $1 
s_{\mu} n s_{e}-$states in helium muonic atoms its contribution equals zero 
identically. 

As mentioned above in any bound state of the helium muonic atom the internal 
muonic orbital is always spherically symmetric, since it is the $1 
s_{\mu}-$orbital. For this orbital the $\langle \delta({\bf r}_{N \mu}) \rangle$ 
expectation value is always a positive, non-zero value. Now, let us consider the 
case of rotationally excited electron's states in helium muonic atoms. In any of 
these states the both electron-nucleus and/or electron-muonic orbitals do not 
have spherical symmetry. Therefore, the both $\langle \delta({\bf r}_{N e}) 
\rangle$ and $\langle \delta({\bf r}_{\mu e}) \rangle$  expectation values equal 
zero identically. For our Eq.(\ref{Htot}) this means that the first two terms 
equal zero, while all other terms are certainly not zero. Furthermore, in the 
last term from Eq.(\ref{Htot}) we always have ${\bf L}_{31} = {\bf L}_{21} = 
{\bf L}$. This allows one to simplify Eq.(\ref{Htot}) even further. Indeed, by 
taking into account all arguments presented here we can write the explicit form 
of the total hyperfine Hamiltonian for the rotationally excited $1 s_{\mu} n 
\ell_{e}-$states in the helium muonic atoms: 
\begin{eqnarray} 
 \hat{H}_{HFS} &=& - \frac{2 \pi \alpha^{2}}{3} \; g_{N} g_{\mu} \; 
 \Bigl(\frac{m_e}{M_p}\Bigr) \; \Bigl(\frac{m_e}{M_{\mu}}\Bigr) \; \langle 
 \delta({\bf r}_{N \mu}) \rangle \; ({\bf I}_{N} \cdot {\bf s}_{\mu}) \nonumber \\  
 &-& \frac14 \; \alpha^{2} \; \Bigl\{ \frac{g_{e} g_N}{M_{p}} \; \Bigl[ 3 \; \langle 
   \; \frac{1}{r^{3}_{31}} \; ({\bf I}_{N} \cdot {\bf n}_{31}) \; ({\bf n}_{31} 
   \cdot {\bf s}_{e}) \; \rangle \; - \; \langle \frac{1}{r^{3}_{31}} \rangle \; 
   ({\bf I}_{N} \cdot {\bf s}_{e}) \Bigr] \nonumber \\ 
 &+& \frac{g_{e} g_{\mu}}{M_{\mu}} \; \Bigl[  3 \; \langle \; \frac{1}{r^{3}_{21}} \; 
  ({\bf s}_{\mu} \cdot {\bf n}_{21}) \; ({\bf n}_{21} \cdot {\bf s}_{e}) \; \rangle 
  \; - \langle \frac{1}{r^{3}_{21}} \rangle \; ({\bf s}_{\mu} \cdot {\bf s}_{e}) \; 
  \Bigr] \Bigr\} \; \nonumber \\ 
 &-& \frac12 \alpha^{2} \; \Bigl[ \frac{g_N}{M_{p}} \; \langle 
 \frac{1}{r^{3}_{31}} \rangle \; ({\bf I}_{N} \cdot {\bf L}) + 
 \frac{g_{\mu}}{M_{\mu}} \; \langle \frac{1}{r^{3}_{21}} \rangle \; 
 ({\bf s}_{\mu} \cdot {\bf L}) \; \Bigr]  \; , \; \label{HtotA} 
\end{eqnarray}
where ${\bf L} = {\bf L}_{31} = {\bf L}_{21}$. Note that the Hamiltonian, 
Eq.(\ref{HtotA}), is the matrix with the dimension $(2 I + 1) (2 s_{\mu} + 1) (2 
s_{e} + 1) (2 L + 1)$ which includes (as coefficients) a few expectation values 
shown in Eq.(\ref{HtotA}) inside of the $\langle \ldots \rangle$ brackets. To 
determine the hyperfine structure of the helium muonic atoms we need to determine 
all eigenvalues (with the corresponding multiplicities) of this matrix. The 
differences between some of these eigenvalues give us the correct numerical values 
of hyperfine structure splittings. 

Thus, the hyperfine structure of the rotationally exited states in helium muonic 
atoms and the corresponding hyperfine structure splittings are determined by our 
formulas  Eq.(\ref{Htot}) and Eq.(\ref{HtotA}) derived above. In fact, these 
formulas can directly be used in numerical calculations of the hyperfine structure 
splittings for an arbitrary rotationally excited (bound) state in the both 
helium muonic ${}^{3}$He $\mu \; e$ and ${}^{4}$He $\mu \; e$ atoms. To achieve 
this goal we need to know the highly accurate wave functions of the rotationally 
excited (bound) states in the helium muonic ${}^{3}$He $\mu \; e$ and ${}^{4}$He 
$\mu \; e$ atoms. In reality, these wave functions are used to determine all 
expectation values mentioned in Eq.(\ref{Htot}) and Eq.(\ref{HtotA}). Possible 
generalizations of our formulas, Eq.(\ref{Htot}) and Eq.(\ref{HtotA}), to the 
case of lithium muonic, beryllium muonic and other similar few-body systems, 
which include one negatively charged muon $\mu^{-}$, is straightforward and here 
we do not want to discuss it. 

In order to conclude this Section let us reduce the formulas Eq.(\ref{Htot}) and 
Eq.(\ref{HtotA}) to the `normal' three-body form. Briefly, this means that all 
tensor terms in these equations must be written as the explicit functions of the 
${\bf n}_{31}$ and ${\bf n}_{32}$ unit vectors only. Right now, each of these 
equations contains the two tensor terms which are included in the second and 
third terms, respectively. However, the first tensor term contains $3 \; \langle 
\; \frac{1}{r^{3}_{31}} \; ({\bf I}_{N} \cdot {\bf n}_{31}) \; ({\bf n}_{31} 
\cdot {\bf s}_{e}) \; \rangle$ tensor term which is already is an explicit 
function of the unit ${\bf n}_{31}$ vector. In contrast with this the second 
tensor term includes a different tensor: $3 \; \langle \; \frac{1}{r^{3}_{21}} 
\; ({\bf s}_{\mu} \cdot {\bf n}_{21}) \; ({\bf n}_{21} \cdot {\bf s}_{e}) \; 
\rangle$ which presents additional problems for our current analysis. Indeed, in 
modern quantum mechanics of three-body systems the origin of coordinates in any 
three-body system usually coincides with the heaviest particle in the system. In 
particular, in the three-body helium muonic atoms the heaviest particle is the 
central helium nucleus, i.e., it is either ${}^{3}$He$^{2+}$, or ${}^{4}$He$^{2+}$ 
nucleus. 

Formally, to correct such a situation we need to represent the unit ${\bf n}_{21}$
vector in the form of linear combination of the unit ${\bf n}_{31}$ and ${\bf 
n}_{32}$ vectors. As is well known for an arbitrary three-body system the following 
equation is obeyed: ${\bf r}_{12} + {\bf r}_{23} + {\bf r}_{31} = 0$. These vectors 
are always coplanar. After a few simple steps this equation is reduced to the form: 
\begin{eqnarray} 
 {\bf n}_{21} = - {\bf n}_{12} = \frac{{\bf r}_{21}}{r_{21}} = \frac{1}{r_{21}} 
 ( {\bf r}_{31} - {\bf r}_{32} ) = \Bigl(\frac{r_{31}}{r_{21}}\Bigr) {\bf n}_{31} 
 - \Bigl(\frac{r_{32}}{r_{21}}\Bigr) {\bf n}_{32} \; \; . \; \; \label{n21} 
\end{eqnarray} 
By using this formula one can derive the following `correct' expression for the 
$({\bf s}_{\mu} \cdot {\bf n}_{21}) \; ({\bf n}_{21} \cdot {\bf s}_{e})$ tensor 
which is included in Eq.(\ref{Htot}) and Eq.(\ref{HtotA}). Here we write this 
expression in the general (or abstract) form:
\begin{eqnarray} 
 ({\bf a} \cdot {\bf n}_{21}) \; ({\bf n}_{21} \cdot {\bf b}) &=&  
 \Bigl(\frac{r_{31}}{r_{21}}\Bigr)^{2} ({\bf a} \cdot {\bf n}_{31}) \; 
 ({\bf n}_{31} \cdot {\bf b}) - \Bigl(\frac{r_{31} r_{32}}{r^{2}_{21}}\Bigr) 
 ({\bf a} \cdot {\bf n}_{31}) \; ({\bf n}_{32} \cdot {\bf b}) \nonumber \\
 &-& \Bigl(\frac{r_{31} r_{32}}{r^{2}_{21}}\Bigr) ({\bf a} \cdot {\bf n}_{32}) 
 \; ({\bf n}_{31} \cdot {\bf b}) + \Bigl(\frac{r_{32}}{r_{21}}\Bigr)^{2} 
 ({\bf a} \cdot {\bf n}_{32}) \; ({\bf n}_{32} \cdot {\bf b}) \; \; . 
 \; \; \label{tensorn21} 
\end{eqnarray} 
where ${\bf a}$ and ${\bf b}$ are the two arbitrary vectors. In the case of 
Eq.(\ref{Htot}) and Eq.(\ref{HtotA}) we have ${\bf a} = {\bf s}_{\mu}$ and 
${\bf b} = {\bf s}_{e}$. Note that the ratio $\Bigl(\frac{r_{31} 
r_{32}}{r^{2}_{21}}\Bigr) \approx \frac{r_{32}}{r_{21}} \approx 
\Bigl(\frac{m_e}{M_{\mu}}\Bigr) \approx \frac{1}{206.77}$ is a small parameter, 
while a similar ratio $\Bigl(\frac{r_{32}}{r_{21}}\Bigr)^{2} \approx 
\Bigl(\frac{m_e}{M_{\mu}}\Bigr)^{2} \approx \frac{1}{(206.77)^{2}}$ is an 
extremely small parameter. These parameters can be used to perform a number of 
non-trivial theoretical evaluations. 

\section{Approximate analysis}

As shown above in any bound state of the helium-muonic atoms we always have 
${\bf L}_{32} = 0$ and also ${\bf L}_{31} = {\bf L}_{21} = {\bf L}$. Formally, 
this means that the original, quite complicated three-body problem is reduced 
to the one-center form. Therefore, we can apply a few well known approximate 
methods, which were  developed earlier for atomic systems with one heavy 
center. In general, the one-center approximation allows one to obtain some 
useful analytical expressions for the tensor terms included in the second term 
of Eq.(\ref{Htot}) and Eq.(\ref{HtotA}) derived above. It is clear that these 
terms cannot be equal zero identically, if (and only if) the nuclear spin 
differs from zero. In respect to this, everywhere below in this Section we 
shall assume that ${\bf I}_{N} \ne 0$ and consider only such one-electron 
muonic atoms/ions. 

In this Section we apply an approximate method which allows one to average 
spatial tensors of any rank over the magnitude of angular momenta $\ell$, 
but not upon its direction is the three-dimensional space, which means that 
below the z-component of angular momentum $\ell_{z}$ is left indeterminate. 
The actual formula which is used here for our present purposes is written 
in the form (see, e.g., \cite{LLQ} and \cite{Fro2025A}):  
\begin{eqnarray}
 \overline{{\bf n}_{i} \; {\bf n}_{j}} = \frac13 \; \delta_{ij} - 
 \frac{1}{(2 \ell - 1) (2 \ell + 3)} \Bigl[ \hat{\ell}_{i} \hat{\ell}_{j} 
 + \hat{\ell}_{j} \hat{\ell}_{i} - \frac23\; \delta_{ij} \; \ell (\ell + 
 1) \Bigr] \; , \; \label{eq9} 
\end{eqnarray} 
where $\hat{\ell}_{i}$ are the Cartesian components of the vector of angular 
momentum of the electron in helium muonic atom, i.e., $\vec{\ell} = ( 
\ell_{x}, \ell_{y}, \ell_{z})$ and ${\bf n}_{i}$ are the Cartesian components 
of an arbitrary unit vector ${\bf n} = (n_x, n_y, n_z)$. As follows from 
Eq.(\ref{Htot}) and Eq.(\ref{HtotA}), where for the helium muonic atoms we 
have the two choices: (1) ${\bf n}_{31} = \frac{{\bf r}_{31}}{r_{31}}$, where 
${\bf r}_{31}$ is the vector directed from the atomic nucleus to the electron, 
and (2) ${\bf n}_{21} = \frac{{\bf r}_{21}}{r_{21}}$, where ${\bf r}_{21}$ is 
the vector directed from the negatively charged muon to the electron. In 
actual three-body calculations the unit vector is represented as a linear 
combination of the unit ${\bf n}_{31}$ and ${\bf n}_{32}$ vectors. This 
problem has been considered in the previous Section. 

Now, let ${\bf a}$ and ${\bf b}$ be two arbitrary vectors, which are 
conserved during time-evolution of our two-body sub-system, then we can 
write Eq.(\ref{eq9}) in a slightly different form  
\begin{eqnarray}
 (\hat{\bf a} \cdot \overline{{\bf n}) ({\bf n}} \cdot \hat{\bf b}) 
 &=& \frac13 (\hat{\bf a} \cdot \hat{\bf b}) - \frac{1}{(2 \ell - 1 ) 
 (2 \ell + 3)} \Bigl[ \hat{a}_{i} \hat{\ell}_{i} \hat{\ell}_{j} 
 \hat{b}_{j} + \hat{a}_{i} \hat{\ell}_{j} \hat{\ell}_{i} \hat{b}_{j} 
 - \frac23\; \; \ell (\ell + 1) \; (\hat{\bf a} \cdot \hat{\bf b}) 
 \Bigr] \; \nonumber \\
 &=& \frac13 (\hat{\bf a} \cdot \hat{\bf b}) - \frac{2}{(2 \ell - 1) 
 (2 \ell + 3)} \Bigl[ (\hat{\bf a} \cdot \hat{\bf \ell}) (\hat{\bf 
 \ell} \cdot \hat{\bf b}) - \frac13\; \; \ell (\ell + 1) \; 
 (\hat{\bf a} \cdot \hat{\bf b}) \Bigr] \; , \; \label{eq91} 
\end{eqnarray} 
where any of these two vectors ${\bf a}$ and ${\bf b}$ cannot be zero, 
or orthogonal to each other. In our case in Eq.(\ref{eq91}) we have to 
choose ${\bf n} = {\bf n}_{31}, {\bf a} = {\bf I}_{N}$ and ${\bf b} = 
{\bf s}_{e}$:
\begin{eqnarray}
 (\hat{\bf I}_{N} \cdot \overline{{\bf n}_{31}) ({\bf n}_{31}} \cdot 
 \hat{\bf s}_e) = \frac13 (\hat{\bf I}_N \cdot \hat{\bf s}_e) - 
 \frac{2}{(2 \ell - 1) (2 \ell + 3)} \Bigl[ (\hat{\bf I}_N \cdot 
 \hat{\bf \ell}) (\hat{\bf \ell} \cdot \hat{\bf s}_e) - \frac13\; \; 
 \ell (\ell + 1) \; (\hat{\bf I}_N \cdot \hat{\bf s}_e) \Bigr] \; 
 . \; \label{eq92} 
\end{eqnarray} 
To obtain the final expression for this term one needs to multiply this 
term by the factor $\frac{3}{r^{3}_{31}}$ and calculate its expectation 
value, i.e., the $\langle \frac{3}{r^{3}_{31}} \rangle$ value. 

\section{Conclusion}

Thus, we developed an original approach which can be used to determine the 
hyperfine structure splittings for the rotationally excited $1 s_{\mu} n 
\ell_{e}$-states in the three-body helium muonic ${}^{3}$He $\mu \; e$ and 
${}^{4}$He $\mu \; e$ atoms. Rigorously, this problem has never been 
considered in earlier studies. Our approach is based on generalization of 
the method proposed by E. Fermi in 1930 \cite{FermiL1} for one-electron 
atomic systems. It is shown that the final formula for the hyperfine 
Hamiltonian $H_{HFS}$ is represented as the sum of three scalar terms, two 
tensor terms and two spin-orbital terms. Diagonalization of the 
corresponding $\hat{H}_{HFS}$ matrix, which has the dimension $(2 I_N + 1) 
(2 s_{\mu} + 1) (2 s_{e} + 1) ( 2 L + 1)$, allows one to determine all 
possible hyperfine structure splittings and investigate the actual 
hyperfine structure in the three-body helium muonic ${}^{3}$He $\mu \; e$ 
and ${}^{4}$He $\mu \; e$ atoms. In reality, we need to determine all 
eigenvalues of the $\hat{H}_{HFS}$ matrix. Differences between these 
eigenvalues (expressed in $MHz$) coincide with the corresponding hyperfine 
structure splittings. All formulas obtained in this study can be used in 
direct numerical calculations of the hyperfine structure splittings in the 
helium muonic atoms. Note also that generalizations of our formulas 
Eq.(\ref{Htot}) and Eq.(\ref{HtotA}) to the cases of lithium muonic, 
beryllium muonic ions and other similar few-body systems with one 
negatively charged muon $\mu^{-}$ is relatively easy, but it requires a 
different system of notation. \\ 

\appendix
 \section{Internal structure of bound states in the helium muonic atoms}
\label{A} 

In this Appendix we describe the internal structure of bound states in 
the three-body helium muonic ${}^{3}$He $\mu \; e$ and ${}^{4}$He $\mu 
\; e$ atoms. This Appendix contains a number of important facts which 
are crucial to understand our evaluations of the basic properties of 
the helium muonic atoms, which are needed to determine the hyperfine 
structure splittings for the rotationally excited bound states. First, 
we have to note that our results for various properties of the bound 
$S(L = 0)-$states in the helium-muonic atoms indicate clearly that all 
these states have the two-shell cluster structure. Indeed, the 
expectation values of muon-nucleus bound state properties are in 
hundreds, thousands and even millions times smaller (or larger) than 
similar electron-nucleus properties (see, e.g., Tables II - IV in 
\cite{Fro2026} and similar Tables from our earlier papers \cite{Fro1}, 
\cite{Fro2}). This means that all bound states in the helium muonic 
${}^{3}$He $\mu \; e$ and ${}^{4}$He $\mu \; e$ atoms have a cluster, 
two-shell structure. 

In other words, the internal structure of each of the bound states in the 
helium muonic atoms is always represented (to very good accuracy) as a  
motion of the negatively charged muon $\mu^{-}$ in the Coulomb field of 
the central positively charged helium nucleus. The electron $e^{-}$ moves 
at large distances form this two-body, very compact, central system (He 
\; $\mu)^{+}$ which can be considered as a heavy, positively charged, 
hydrogen-like quasi-nucleus with the unit electric charge. The electric 
charge of this central quasi-nucleus equals $e$ (or $+ e$). This heavy 
quasi-nucleus can attract one, or two negatively charged electrons and 
form either the three-body helium muonic He$^{2+} \; \mu^{-} \; e^{-}$ 
atom, or the four-electron, negatively charged He$^{2+} \; \mu^{-} \; 
e^{-}_{2}$ ion. The last ion is very similar to the two-electron, 
negatively charged H$^{-}$ ion, which is the Coulomb three-body system. 
Based on this analogy we can predict that each of the four-body helium 
muonic  ${}^{3}$He $\mu \; e_2$ and ${}^{4}$He $\mu \; e_2$ ions has 
only one bound (ground) $1^{1}S_{e}-$state.     

Briefly, we can say that each of the helium muonic atoms and/or negatively 
charged ions has the two shells: muon shell (or internal shell) and 
electron shell (or outer-most shell). Spatial radius of the muonic shell 
is in $\Bigl( \frac{m_{\mu} Q}{m_{e}}\Bigr) \approx$ 413.54 times smaller 
than analogous radius of the electron shell. Such a two-shell 
approximation, which has been proposed in 1996 and extensively used in 
all our later studies, is based on this fundamental fact. In general, 
any bound state in the three-body (neutral) helium-muonic He $ \mu \; 
e$ atom always contains two shells: (a) muonic (or internal) shell, and 
(b) electron (or outer-most) shell. The muonic shell radius is more than 
$\Bigl( \frac{m_{\mu} Q}{m_{e}}\Bigr) \approx$ 413.54 times smaller than 
the electron's shell radius. Our two-shell approximation is based on this 
obvious and fundamental fact.

Note that the spatial radius of the internal (compact) two-body (He 
$\mu)^{+}$ ion is $R_{\mu} \approx \frac{a_{0}}{2 m_{\mu}} < \Lambda_{e} 
= \alpha a_{0}$, where $\Lambda_{e} = \frac{\hbar}{m_e c}$ is the reduced 
(electron's) Compton wavelength. The spatial radius of the (He $ \mu)^{+} 
\; e^{-}$ atom equals $\approx a_{0}$, where $a_{0} = \frac{\hbar^{2}}{m_e 
e^{2}}$ is the Bohr radius. The corresponding wave function is represented 
as a direct product of the two wave functions, where the first wave 
function describes the internal (or central), two-body helium muonic (He 
$\mu)^{+}$ ion, while the second wave function corresponds to the regular 
one-electron hydrogen atom. Again this means that in any helium muonic 
atom the (He $\mu)^{+}$ ion plays the role of a central heavy nucleus (or 
quasi-nucleus). 

Let us consider this problem from a different point of view. As is well known 
the total symmetry of the Schr\"{o}dinger equation for one-electron 
hydrogen-like atoms/ions is described by the compact $SO(4)$ group \cite{Pauli} 
(see, also \cite{BarR} and \cite{Fro2025A}). In respect to this, the spinless 
wave functions of an arbitrary bound state in hydrogen-like atoms and/or ions 
can be designated by some three-dimensional vector $\mid n, \ell, m \rangle$ 
\cite{Fro2025A}, where $n, \ell$ and $m$ are the integer numbers. In reality, 
the three components of this vector are the three conserving (integer) quantum 
numbers which are: (a) the principal quantum number $n (\ge 1)$, (b) the 
orbital quantum number $\ell (\ge 0)$, and (c) magnetic quantum number $m$. 
These three quantum numbers obey the following constraints: $n \ge \ell + 1, 
\ell \ge 0$ and $-\ell \le m \le \ell$.    

Based on these arguments we can assume that classification of bound states in 
helium-muonic atoms can be made in terms of the compact $SO(4) \otimes SO(4)$ 
group which describes (almost exactly) the actual symmetry of each of these 
two-shell cluster systems. Briefly, this means that an arbitrary bound state 
in the three-body helium muonic (He $\mu)^{+} e^{-}$ atom can be denoted as 
the six-dimensional vector of integer numbers: $\mid n_{\mu} \ell_{\mu} 
m_{\mu} \rangle \mid n_{e} \ell_{e} m_{e} \rangle$, where $n_{\mu}$ and 
$n_{e}$ are the principal quantum numbers of muon and electron, respectively. 
Analogously, $\ell_{\mu}$ and $\ell_{e}$ are the muon and electron angular 
momenta, respectively, while $m_{\mu}$ and $m_{e}$ are the corresponding 
magnetic quantum numbers. This classification scheme works perfectly for the 
bound states in the helium muonic atoms, lithium and beryllium muonic ions 
and other similar systems. 

In reality, in any bound state of the three-body helium muonic atoms we 
always have $n_{\mu}  = 1, \ell_{\mu} = 0$ and $m_{\mu} = 0$, since all 
states with $n_{\mu} > 1$ are unstable against radiation-free 
self-ionization 
\begin{eqnarray}
  He \; \mu \; e \; \; \; ({\rm state \; with} \; n_{\mu} > 1) = 
  ( He \; \mu)^{+} \; \; \; ({\rm state \; with} \; n_{\mu} = 1) \; 
  + \; e^{-} \; \; , \; \label{A1}
\end{eqnarray}
where the free electron removes all excess of energy which is released 
during this process. The probability of this Auger process, Eq.(\ref{A1}), 
(or internal conversion) is relatively large, since it does not include 
the fine-structure constant. This means that all rotationally excited 
(bound) states in the helium muonic atoms is always represented by the 
formula $\mid 1_{\mu} 0_{\mu} 0_{\mu} \rangle \mid n_{e} \ell_{e} 0 
\rangle$, where $n_e \ge 2$ and  $\ell_e \ge 1$. In some cases it is 
better to designate the values of $\ell$ by the small letters which 
corresponds to the well known spectroscopic order, where the letter 
$s$ stands for $\ell = 0$, while other letters $p, d, f, \ldots$, 
designate the following values of $\ell$: $\ell = 1, 2, 3, \ldots$.


\begin{thebibliography}{99}

\bibitem{Dir28} P.A.M. Dirac, Proc. Roy. Soc. {\bf A117}, 610 (1928). 

\bibitem{FermiL1} E. Fermi, Zeit. f\"{u}r Physik {\bf 60}, 320 (1930) 
(see, also his short note in: Nature {\bf 125}, 16 (1930)). 

\bibitem{Sobel} I.I. Sobel'man \textit{An Introduction to the Theory of Atomic 
Spectra} (Pergamon Press, N.Y. (2014)), Chapt. 6.  

\bibitem{Heine} V.H. Heine, \textit{Group Theory in Quantum Mechanics} , (Dover 
Publ. Inc., Mineola, New York, 1990), Chapt. IV. 

\bibitem{Hughes} V.W. Hughes, G.z. Putliz, in: T. Kinoshita (Ed.), 
\textit{Quantum Electrodynamics}, (World Scientific Publishing Co. Pte. Ltd., 1
990), Chapter 16.

\bibitem{Huan1} K.N. Huang and V.W. Hughes, Phys. Rev. A {\bf 20}, 706 (1979). 

\bibitem{Huan2} K.N. Huang and V.W. Hughes, Phys. Rev. A {\bf 26}, 2330 (1982).

\bibitem{Lakd} S.D. Lakdawala and P.J. Mohr, Phys. Rev. A {\bf 22}, 1572 (1980).

\bibitem{Lakd1} S.D. Lakdawala and P.J. Mohr, Phys. Rev. A {\bf 24}, 2224 (1981).

\bibitem{Drachm1} R.J. Drachman, J. Phys. B {\bf 14}, 2733 (1981).

\bibitem{Drachm2} R.J. Drachman, J. Phys. B {\bf 16}, L479 (1983).

\bibitem{Chen} M.K. Chen and C.S. Hsue, Phys. Rev. A {\bf 22}, 3951 (1989). 

\bibitem{Fro1} A.M. Frolov, Phys. Rev. A {\bf 61}, 022509 (2000). 

\bibitem{Fro2} A.M. Frolov, Phys. Rev. A {\bf 65}, 024701 (2002). 


\bibitem{Fro2026} A.M. Frolov, European Phys. Journal D {\bf 80}, article \# 89 (2026), 
(Article DOI: /10.1140/epjd/s10053-026-01192-1).  

\bibitem{NIST} see, e.g., https://physics.nist.gov/cgi-bin/cuu/Value? 

\bibitem{Jackson} J.D. Jackson, \textit{Classical Electrodynamics}, (2nd ed., John 
Willey \& Sons (1975)), Sect. 5.8. 

\bibitem{LLQ} L.D. Landau and E.M. Lifshitz, {\it Quantum Mechanics: Non-Relativistic 
Theory}, (3rd. ed. Pergamon Press, New York (1976)), \$ 121. 

\bibitem{Pauli} W. Pauli, Zeit. f\"{u}r Physik {\bf 36}, 336 (1926). 

\bibitem{BarR} A.O. Barut and R. Razcka, \textit{Theory of Group Representations and Applications} 
(World Scientific, Singapore (1986)), Chapters 8 - 11. 

\bibitem{Fro2025A} A.M. Frolov, European Phys. Journal D {\bf 79}, article \# 13 (2025), 
(Article DOI: 10.1140/epjd/s10053-025-00966-3).

\end{thebibliography}
\end{document}